\documentclass[a4paper]{article}

\usepackage{graphicx}
\usepackage{float}
\usepackage{epstopdf}

\newcommand{\diff}{\mathrm{d}}

\title{Silhouette and spectral line profiles in the special modification of the Kerr black hole geometry generated by quintessential fields}
\author{Jan Schee\footnote{jan.schee@fpf.slu.cz} and Zden\v{e}k Stuchl\'{i}k\footnote{zdenek.stuchlik@fpf.slu.cz}\\
	 \small{Institute of Physics and Research Centre for Theoretical Physics \& Astrophysics}\\
	 \small{Faculty of Philosophy and Science}\\
	 \small{Silesian university in Opava}\\
	 \small{Bezru\v{c}ovo n\'{a}m. 13, 74601 Opava}\\
	 \small{Czech Republic}}
\date{}

\begin{document}

\maketitle

\begin{abstract}
We study optical effects in quintessential Kerr black hole spacetimes corresponding to the limiting case of the equation-of-state parameter $\omega_{q}=-1/3$ of the quintessence. In dependence on the dimensionless quintessential field parameter $c$, we determine the black hole silhouette and the spectral line profiles of Keplerian disks generated in this special quintessential Kerr geometry, representing an extension of the general modifications of the Kerr geometry introduced recently by Ghasemi-Nodehi and Bambi \cite{Gha-Bam:2016:EPJC:}. We demonstrate that due to the influence of the parameter $c$, the silhouette is almost homogeneously enlarged, and the spectral line profiles are redshifted with almost conserved shape.   
\end{abstract}

\section*{Introduction}

Recent research of astrophysical and connected optical phenomena related to possible direct observations of effects occuring in the field of black holes and naked singularities, or other strong gravity compact objects, is often concentrated in effects occuring in the modified Kerr geometry implied by alternatives to the Einstein gravity. Observable effects related to these modified Kerr geometries could give signatures of the alternatives or extensions of the standard Einstein gravity, or of the influence of non-vacuum environment of the compact objects. The modifications of the Kerr geometry can be separated into two classes. 

In the first class, modifications of the Kerr (or Schwarschild) geometry are related to a concrete modification of the standard Einstein gravity. There is a large variety of research related to the braneworld models \cite{Dad-etal:2000:PhysLetB:,Ali-Gum:2005:PHYSR4:,Kot-Stu-Tor:2008:CLAQG:,Stu-Kot:2009:GRG:,Sche-Stu:2009:GRG:,Sche-Stu:2009:IJMPD:,Abd-Ahm:2010:PHYSR4:,Ama-Eir:2012:PHYSR4:,Sha-Ahm-Abd:2013:PHYSR4:,Ova-Ger-Cas:2015:CLAQG:}, standard gravity combined with the non-linear electrodynamics giving regular black holes or no-horizon spacetimes \cite{Ayo-Gar:1998:PhysRevLet:,Ayo-Gar:1999:GRG:,Bam-Mod:2013:PhysLetB:,Stu-Sche:2015:IJMPD:,Sche-Stu:2015:JCAP:,Tos-etal:2014:PHYSR4:}, Ho\v{r}ava quantum gravity \cite{Hor:2009:PHYSR4:,Keh-Sfe:2009:PhysLetB:,Vie-etal:2014:PHYSR4:,Stu-Sche:2014:CLAQG:,Stu-Sche-Abd:2014:PHYSR4:}, String theory \cite{Sen:1992:PHYSRL:}, f(R) gravity \cite{Sot-Far:2010:RevModPhys:,Per-Rom-Per:2013:ASTRA:}, Kerr geometry modified by the cosmological constant \cite{Car:1973:BlaHol:,Stu:1983:BAC:,Stu-Cal:1991:GRG:,Stu-Sla-Hle:2000:ASTRA:,Rez-Zan-Fon:2003:ASTRA:,Stu-Sla:2004:PHYSR4:,Sla-Stu:2005:CLAQG:,Stu:2005:MPLA:,Kra:2005:CLAQG:,Kra:2007:CLAQG:,Kag-etal:2006:PhysLetB:,Stu-Kov:2008:IJMPD:,Stu-Sla-Kov:2009:CLAQG:,Hac-etal:2010:PHYSR4:,Hac-etal:2010:PHYSR4:,Oli-etal:2011:MPLA:,Stu-Sche:2011:JCAP:,Kra:2011:CLAQG:,Kra:2014:GRG:}, or quintessential fields \cite{Kis:2003:CLAQG:,Fer-Med-Rei:2015:IJTP:,Tos-Stu-Ahm:2015:ArXiv:}, higher-dimensional spacetimes \cite{Hac-etal:2010:PHYSR4:}. 

In the second class, the Kerr geometry is modified in the framework of the so called dirty Kerr geometry \cite{Joh-Psa:2011:PHYSR4:,Car-Pan-Ric:PhRD:2014:}, where the modifications are usually unrelated to any source term. The Kerr (Schwarzschild) vacuum solution of the Einstein equations is considered as an etalon of these studies, and its metric coefficients are correspondingly modified. In the dirty Kerr geometry framework, the radial profiles of the metric coefficients are modified and effects of the modified metric on the optical phenomena are studied \cite{Bam:2013:JCAP:,Bam:2015:CLAQG:,Jia-Bam-Ste:2015:JCAP:}. Recently, Ghasemi-Nodehi and Bambi proposed a new version of modified Kerr geometry, where also the latitudinal profiles of the metric coefficients can be modified \cite{Gha-Bam:2016:EPJC:}. In this approach, the geometry modifications are related to all terms of the Kerr metric containing mass $M$ and specific angular momentum (spin) $a$. However, there could exist also modifications of the Kerr metric that are not related to the parameters $M$ and $a$. Here we briefly study a concrete case of such a spacetime, related to the quintessential rotating black holes. 

Here we concentrate attention to a special class of the quintessential Kerr black holes generating modification of the Kerr geometry that represents an extension of the seemingly most general class of modifications of the Kerr geometry introduced in \cite{Gha-Bam:2016:EPJC:}. The considered modification of the Kerr geometry occurs due to a special (limit) class of the quintessential field. 

Recently, Toshmatov, Stuchlik and Ahmedov \cite{Tos-Stu-Ahm:2015:ArXiv:} have found an axially symmetric stationary non-vacuum solution of the Einstein equations where the energy-momentum tensor corresponds to the quintessential scalar fields. The quintessential field is characterized by the equation-of-state parameter $\omega_{q} \in (-1,-1/3)$, and by the dimensionless parameter $c$ representing intensity of the quintessential field. The quintessential Kerr (Schwarzschild) geometry demonstrates repulsive gravity at large distances, and implies existence of the cosmological horizon \cite{Tos-Stu-Ahm:2015:ArXiv:}. However, in the limit case of $\omega_{q} \to -1/3$, the quintessential Kerr geometry becomes the asymptotically flat Kerr geometry - the geometry is modified by the quintessential field parameter $c$, but the dependence of the metric coefficients on the mass and spin parameters, $M,a$, remains to be of the standard Kerr character. 

Here we study the black hole silhouette shape, and the profile of the spectral lines of radiation from the inner parts of the Keplerian disks for the special limit ($\omega_{q}=-1/3$) of the quintessential Kerr metric, in dependence on the parameter $c$. Both these phenomena are fully governed by the geodesic structure of the quintessential Kerr spacetime. 
 
\section{The quintessential Kerr spacetimes and their special limit}

The geometry of the quintessential rotation black hole spacetime introduced in \cite{Tos-Stu-Ahm:2015:ArXiv:} is described in the standard Boyer-Lindquist coordinates and the geometric units with $c=G=1$ by the line element
\begin{equation}
ds^2=g_{tt} dt^2+g_{rr} dr^2+2g_{t\phi} drd\phi+g_{\theta\theta}d\theta^2+g_{\phi\phi}d\phi^2
\end{equation}
with
$$g_{tt}=-1+\frac{2Mr+cr^{1-3\omega_q}}{r^2+a^2\cos^2{\theta}},$$
$$g_{rr}=\frac{r^2+a^2\cos^2{\theta}}{a^2-2Mr+r^2-cr^{1-3\omega_q}},$$
$$g_{t\phi}=-a\sin^2{\theta}\frac{2Mr+cr^{1-3\omega_q }}{r^2+a^2 \cos^2{\theta}},$$
$$g_{\theta\theta}=r^2+a^2\cos^2{\theta},$$
$$g_{\phi\phi}=\sin^2{\theta}(a^2+r^2+a^2 \sin^2{\theta}\frac{2Mr+cr^{1-3\omega_q}}{r^2+a^2 \cos^2{\theta}})$$
where $M$ and $a$ are the gravitational mass and the specific angular momentum $a=J/M$ of the black hole, $\omega_{q} \in (-1/3,-1)$ is the quintessential scalar field equation-of-state parameter, while $c \in (0,1)$ is the quintessential field parameter characterizing the field magnitude. When the quintessential field parameter $c=0$, the rotating quintessential geometry reduces to the Kerr geometry. The related quintessential field energy-stress tensor is also determined in \cite{Tos-Stu-Ahm:2015:ArXiv:}, but only the spacetime geometry will be relevant in our study. \footnote{This quintessential Kerr spacetime has been obtained as a generalization of the Kisilev solution for static quintessential black holes \cite{Kis:2003:CLAQG:}, using the standard method developed by Newman and Janis \cite{New-Jan:1965:JMathPhys:}, and those presented in  \cite{Aze-Ani:2014:PHYSR4:}. An alternative based on the method presented in \cite{Bam-Mod:2013:PhysLetB:,Tos-etal:2014:PHYSR4:} has been introduced in \cite{Gho:2016:EPJC:}.} 

In the case of the special limit of the quintessential Kerr black holes, determined by the equation of state parameter $\omega_{q}=-1/3$, the line element of the geometry simplifies to the form 
\begin{eqnarray}\label{line_element}
	\diff s^2 &=& -\left(1-\frac{2\rho r}{\Sigma}\right)\diff t^2 + \frac{\Sigma}{\Delta}\diff r^2+\Sigma\diff\theta^2\nonumber\\
	&&-\frac{4 a \rho r\sin^2\theta}{\Sigma}\diff t\diff\phi + \left(r^2+a^2+a^2\sin^2\theta\frac{2\rho r}{\Sigma}\right)\sin^2\theta\diff\phi^2
\end{eqnarray}	
where 
\begin{eqnarray}
	\Sigma&=&r^2+a^2\cos^2\theta,\\
	\Delta&=&r^2-2\rho r+a^2,\\
	2\rho(r) &=& 2+c r.
\end{eqnarray}
and the mass parameter is put for simplicity to $M=1$, i.e., we express the spin and quintessential field parameter $c$ as dimensionless quantities, and we use dimensionless radial and time coordinates $r/M \to r, t/M \to t$. The special ($\omega_{q}=-1/3$) quintessential Kerr metric clearly represents an extension of the Ghasemi-Nodehi--Bambi modification of the Kerr geometry, as it modifies also the $r^2$ term in the expression $\Delta = r^2 - 2Mr + a^2$ governing the loci of the black hole event horizons. In contrast to the quintessential black hole spacetimes with $\omega_{q} \in (-1,-1/3)$, having a cosmological horizon, the special case of the $\omega_{q}=-1/3$ spacetimes is asymptotically flat. The horizons are given by the relation 
\begin{eqnarray}
	r_{\pm} = \frac{1 \pm \sqrt{1-a^2(1-c)}}{1-c}.
\end{eqnarray}
Now the limit on the black hole spin reads
\begin{eqnarray}
	a^2 \leq \frac{1}{1-c} , 
\end{eqnarray}
and it breaks the standard black hole bound of $a \leq 1$. 

We study simple optical effects in this special quintessential rotating black hole geometry in order to demonstrate the role of this kind of the Kerr geometry modifications. It is convenient in numerical simulations to use the reciprocal radial coordinate $u\equiv 1/r$ and the latitudinal coordinate $m\equiv\cos\theta$. The line element (\ref{line_element}) then takes the form 
\begin{eqnarray}
	\diff s^2&=&-\left(1-\frac{2\tilde{\rho}}{\tilde{\Sigma}u}\right)\diff t^2+\frac{\tilde{\sigma}}{\tilde{\Delta}u^3}\diff u^2-\frac{4a\tilde{\rho}(1-m^2)}{u\tilde{\Sigma}}\diff\phi\diff t\nonumber\\
			&&+\frac{\tilde{\Sigma}}{u^2(1-m^2)}\diff m^2+\frac{1-m^2}{u^2}\left[1+a^2u^2+\frac{2a^2(1-m^2)u^2\tilde{\rho}}{\tilde{\sigma}}\right]\diff\phi^2,
\end{eqnarray} 
where is
\begin{eqnarray}
	\tilde{\Sigma}&=&1+a^2u^2 m^2,\\
	\tilde{\Delta}&=&u-2\tilde{\rho}+a^2u^3,\\
	2\tilde{\rho}&=&cu+2u^2.
\end{eqnarray}

\section{Equations of geodesic motion}

The equations of geodesic motion are given by the Hamiltonian \cite{Car:1973:BlaHol:}
\begin{equation}
	H=\frac{1}{2}g^{\mu\nu}p_{\mu} p_{\nu}
\end{equation}
where $H=0$ for null-geodesics, and $H=-\frac{1}{2}m^2$ for massive test particles; and $p_{\mu}$ are covariant components of the test particle 4-momentum. There are two constants of the motion related to the spacetime symmetries: energy $E$, connected with the stationarity, axial angular momentum $L$, connected with the axial symmetry.  

For the special quintessential rotating black hole geometry, given by the line element (\ref{line_element}), we obtain the set of first order differential equations \cite{Tos-Stu-Ahm:2015:ArXiv:} 
\begin{eqnarray}
\Sigma p^t&=&\frac{r^2+a^2}{\Delta}\left[E(r^2+a^2)-aL\right]-a\left(aE\sin^2\theta-L\right),\\
\Sigma p^r&=\pm\sqrt{R},\\
\Sigma p^\theta&=\pm\sqrt{W},\\
\Sigma p^\phi&=&\frac{a}{\Delta}\left[E(r^2+a^2)-aL\right]-\left(aE-\frac{L}{\sin^2\theta}\right),
\end{eqnarray}
where 
\begin{eqnarray}
	R(r;l,q,a,c)&=&\left[(r^2+a^2)E-aL\right]^2-\Delta\left[(aE-L)^2+mr^2+Q\right],\\
	W(\theta;l,q,a,c)&=&Q-\left[\frac{L^2}{\sin^2\theta}+a^2(m^2-E^2)\right]\cos^2\theta;
\end{eqnarray}
$Q$ is the standard fourth integral of the motion in the Kerr (and modified Kerr) spacetimes -- it reflects hidden symmetries of the spacetime and is related to the total angular momentum of the motion. Notice that the latitudinal motion is governed by the same equation as in the Kerr geometry, but the equation of the radial motion is modified due to the presence of the quintessential field. 

\section{Keplerian orbits}

\begin{figure}[H]
\begin{center}
\begin{tabular}{cc}
	\includegraphics[scale=0.5]{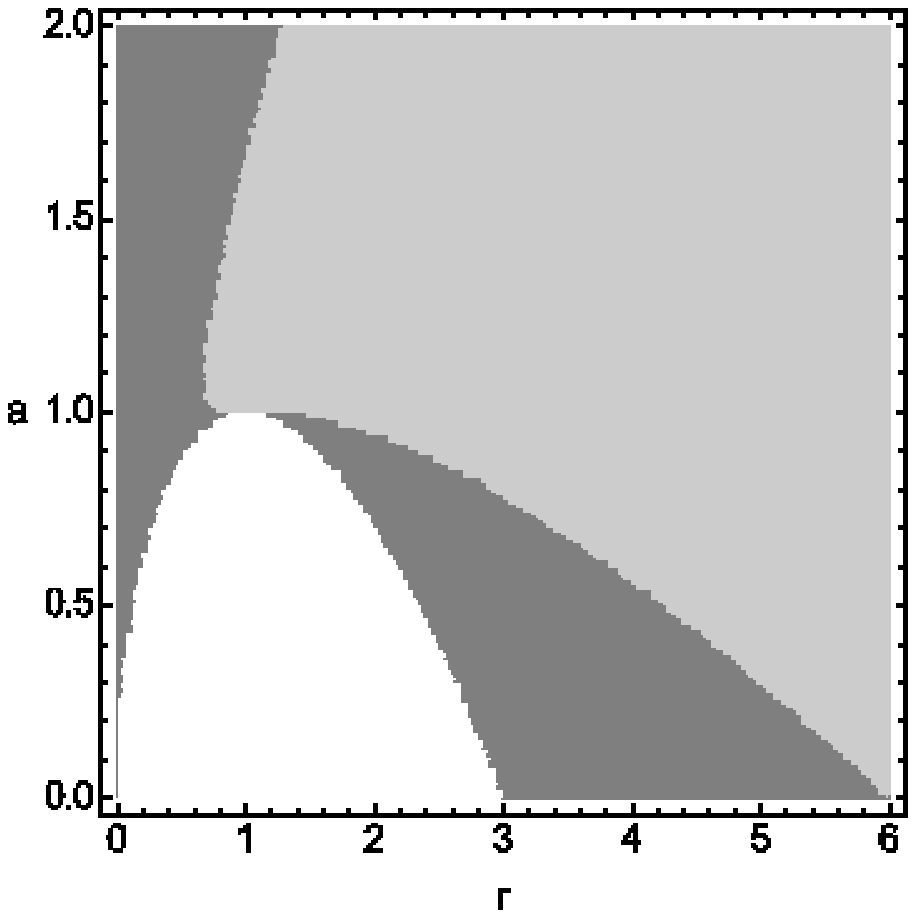}&\includegraphics[scale=0.5]{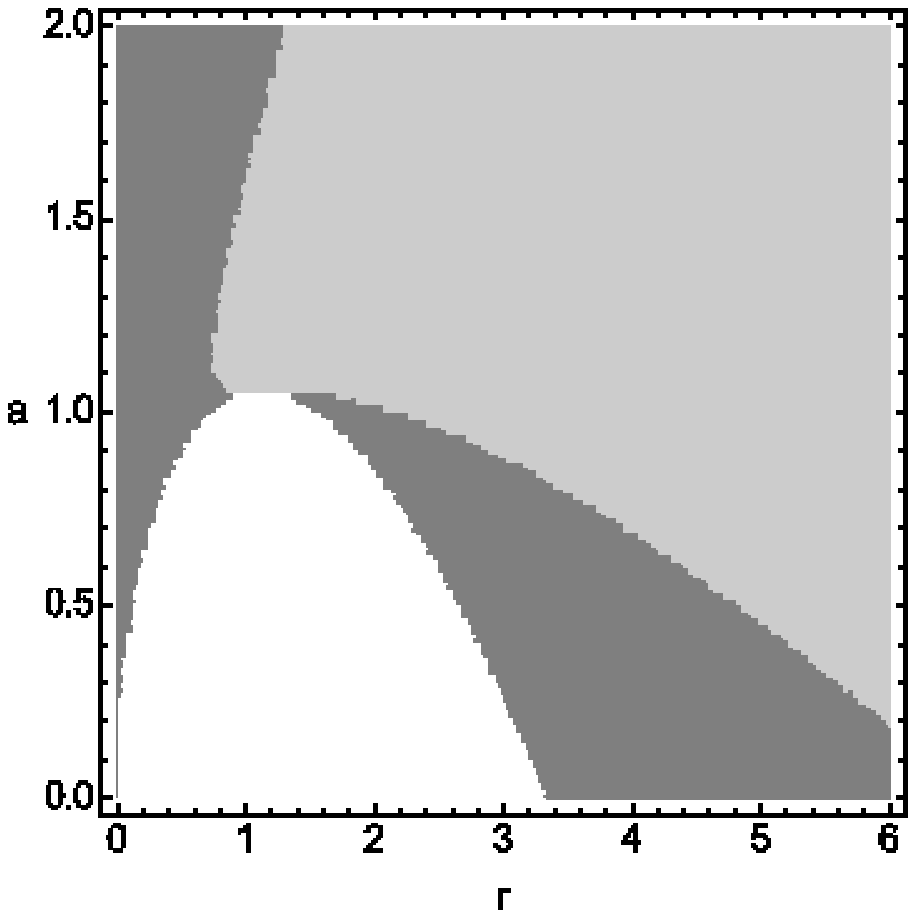}
\end{tabular}
\caption{\label{circ-orm-map}The $(a,r)$ map of circular orbits in the quintessential rotating spacetime for two representative values of parameter $c=0$, and $10^{-1}$ (left to right). Non circular orbits are allowed in the white regions. The stable circular orbits lie with within the gray regions, and unstable circular orbits fall into dark gray regions. }
\end{center}
\end{figure}

We illustrate the effect of the special quintessential rotating black hole geometry on the profiled spectral line of the radiation originating in the Keplerian disks assumed to be composed of the test particle emitters following the stable circular orbits with so called Keplerian angular frequency. Therefore, we first discuss the circular geodesics and their stability. The circular geodesics are situated in the equatorial plane, having $Q=0$, as in the case of the Kerr spacetimes \cite{Bar:1973:BlaHol:,Stu:1980:BAC:}. The resulting maps of the circular geodesics are summarized in Fig. \ref{circ-orm-map}. 

\begin{figure}[H]
	\begin{center}
	\begin{tabular}{cc}
		\includegraphics[scale=0.5]{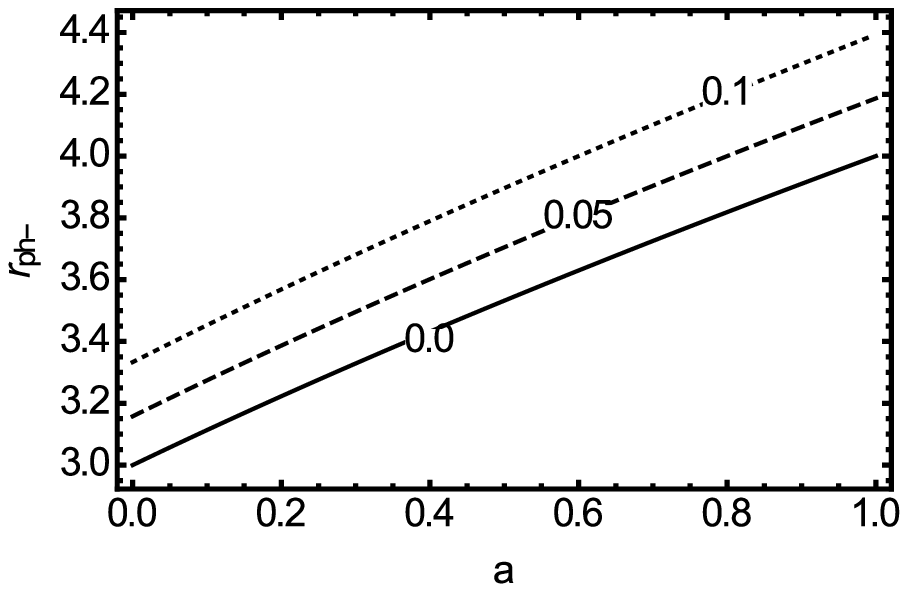}&\includegraphics[scale=0.5]{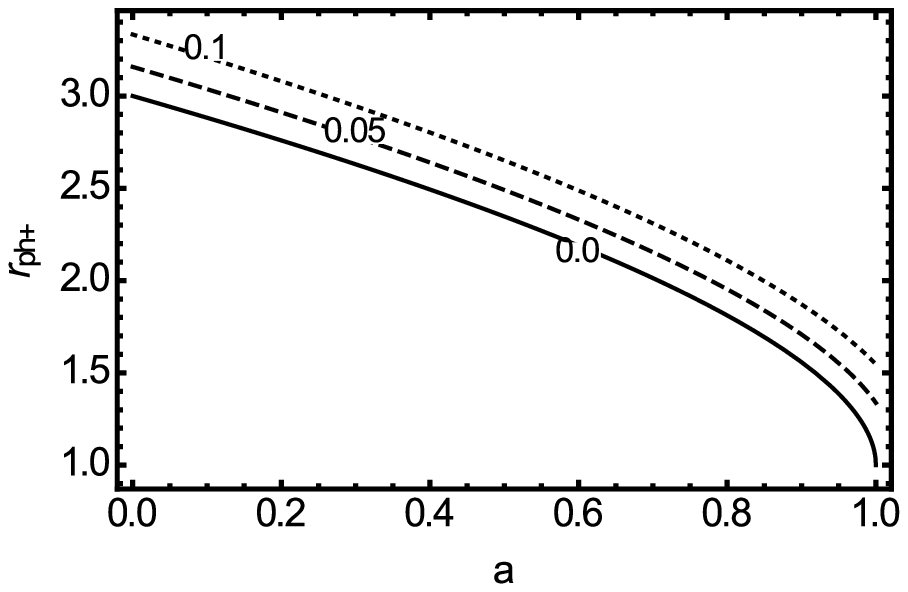}\\
		\includegraphics[scale=0.5]{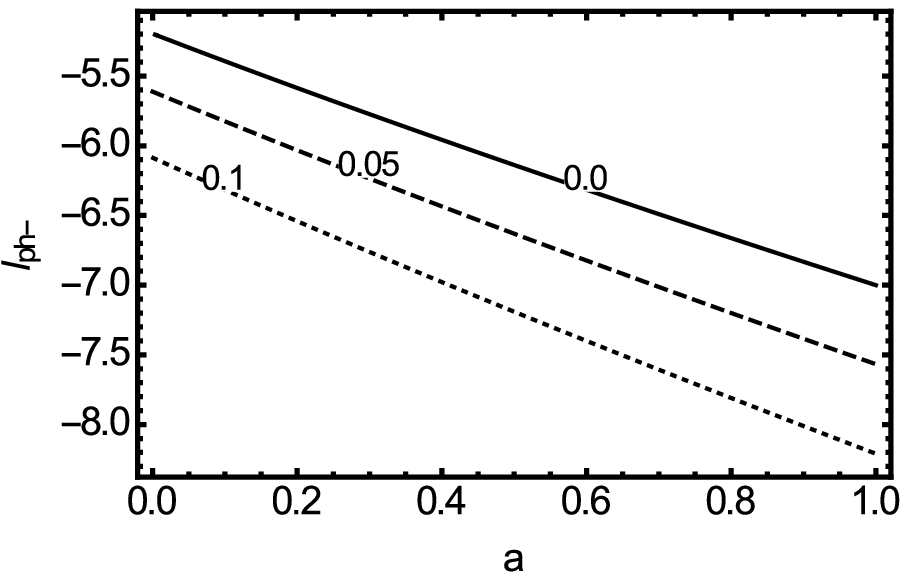}&\includegraphics[scale=0.5]{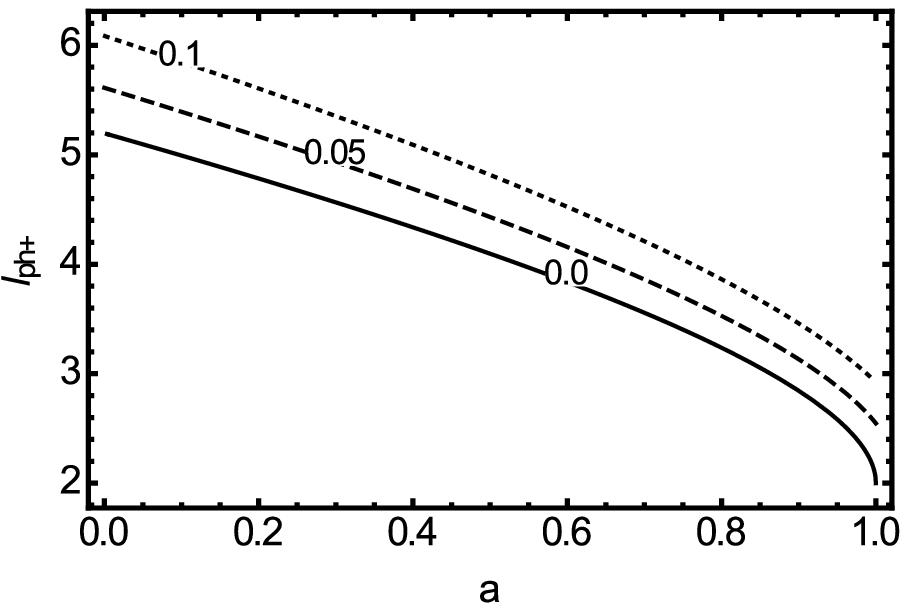}
	\end{tabular}
	\caption{\label{photon-circ-orb}Radii and corresponding impact parameters of both  co-rotating(right) and counter-rotating (left) photon circular orbits plotted as functions of spin parameter $a$ for three representative values of parameter $c=0.0$, $0.05$, and $0.1$. } 
	\end{center}
\end{figure}

The circular test particle orbits are determined by the conditions
\begin{equation}
	R(r;E,L,a,c)=0 , \frac{\diff R(r;E,L,a,c)}{\diff r}=0 
\end{equation} 
that have to be satisfied simultaneously. The radial profiles of the specific energy $E/m$ and the specific axial angular momentum $L/m$ take the following form 
\begin{equation}
	\frac{E_\pm^2(r,a,c)}{m} = \frac{-r^3(2+\sigma r)^2(3+\sigma r)+a^2r^2(5+3\sigma r)\pm 2\Delta a r^{3/2}}{r^3[-4a^2+r(3+\sigma r)^2]}
\end{equation} 
and
\begin{eqnarray}
	\frac{L_\pm^2(r,a,c)}{m} &=& \frac{[-r^3(2+\sigma r)^2(3+\sigma r)+a^2r^2(5+3\sigma r)\pm 2\Delta a r^{3/2}]}{a^2\Delta^2 r^3(-4a^2+r(3+\sigma r)^2)}\nonumber\\
		&\times& [a^4+a\Delta r^{3/2}+a^2r(4+r(-3+4c+\sigma c r))]^2
\end{eqnarray} 
where we have introduced $\sigma\equiv -1+c$. The limits on the existence of the circular geodesics are given by the photon circular geodesics. Their loci are the roots of the equation
\begin{eqnarray}
	0&=&\sigma^4r^6+8\sigma^3r^5+2(11-8a^2)\sigma^2 r^4-4\sigma(-6+a^2(7+c))r^3+[9+16a^4\sigma-16a^2c]r^2\nonumber\\
	&&+4a^2(-3+4a^2)r-4a^2(-1+a^2).
\end{eqnarray} 
while their impact parameters are given by 
\begin{equation}
	l^2_{ph\pm} \equiv\frac{L}{E} = \frac{a(1+cr)\pm r\sqrt{a^2-2r(1+\sigma r)}}{1+\sigma r}.
\end{equation} 
We plot the $r_{ph\pm}$ and $l_{ph\pm}$ as function of spin $a$ for three representative values of parameter $c=0$, $0.05$, and $0.1$ in Fig. \ref{photon-circ-orb}
The stability of the circular geodesics is encoded in the sign of $\diff^2 R/\diff r^2$ under the conditions related to the circular orbits: 
\begin{eqnarray}
		\frac{\diff^2 R}{\diff r^2}>0&\Rightarrow& \textrm{ unstable orbit}\\
		\frac{\diff^2 R}{\diff r^2}<0&\Rightarrow& \textrm{ stable orbit}\\
		\frac{\diff^2 R}{\diff r^2}=0&\Rightarrow& \textrm{ marginally stable orbit}
\end{eqnarray}

In Fig. \ref{circ-orm-map} we demonstrate dependence of the location of the unstable and stable circular geodesics on the black hole spin $a$ for the characteristic values of the quintessential field parameter $\tilde{q}$. 

\section{Silhouette of the special quintessential rotating black holes}

\begin{figure}[H]
	\begin{center}
		\includegraphics[scale=0.7]{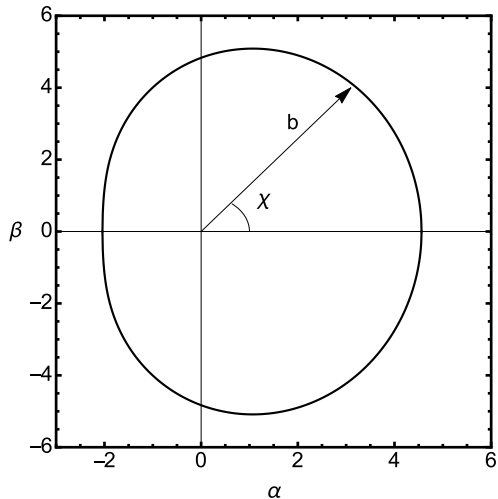}
		\caption{\label{def-b-chi}Definition of coordinates $b$ and $\chi$ to characterize the silhouette shape.}
	\end{center}
\end{figure}

The black hole silhouette is intimately connected with the photon spherical orbits for which the photon 4-momentum $k^{\mu}$ satisfies the conditions \cite{Bar:1973:BlaHol:} 
\begin{equation}
	k^u=0 , \frac{\diff k^u}{\diff\lambda}=0
\end{equation}
implying equations relating the impact parameters $l=L/E$ and $q=Q/E^2$
\begin{equation}
	R(r,l,q;a,c) =0,  \frac{\diff R(r,l,q;a,c)}{\diff r}=0.\label{zero}
\end{equation}
Expressing $q\equiv q(r,l;a,c)$ from the equation (\ref{zero}), we obtain the conditions for the spherical photon orbits in the form 
\begin{eqnarray}
	q(r;l,a,c)&\equiv& \frac{(r^2+a^2-a l)^2}{\Delta(r;a,c)}-(a-l)^2,\\
	\frac{\diff q(r;l,a,c)}{\diff r}&=&\frac{1}{\Delta^2}\left[4r(r^2+a^2-a l)\Delta-(r^2+a^2-      al)\Delta'\right]=0.\label{eqph}
\end{eqnarray}
The silhouette is constructed for a static distant observer with a detector equipped with the coordinates $(\alpha,\beta)$ which are connected with the impact parameters $(l,q)$ through relations
\begin{equation}
	l=-\alpha/\sqrt{1-m_o^2},\textrm{ and } q = \beta^2+ m_o^2(\alpha^2-a^2).
\end{equation}
For fixed parameters $a$, $c$ and $l\in[l_{min}, l_{max}]$, we can determines the corresponding radius of the photon spherical orbits $r_{sp}>r_+$.
The values of $l_{min}$ and $l_{max}$ follow from the condition $\beta=0$ which implies $q=(l^2+a^2)$ for $\theta_o=\pi/2$. Then we solve the equations
\begin{equation}
	\Delta (l^2-a^2)=(r^2+a^2-a l)^2-\Delta(a-l)^2
\end{equation}
and (\ref{eqph}) simultaneously to obtain $(r_{ph-min},l_{min})$ and $r_{ph-max},l_{max}$. It is also useful to transform the silhouette coordinates $(\alpha,\beta)$ to the new coordinates $(b,\chi)$ defined as (see Fig. \ref{def-b-chi})
\begin{equation}
	b^2=\alpha^2+\beta^2, \quad \chi=\arctan\left(\frac{\beta}{\alpha}\right).
\end{equation}

\begin{figure}[H]
 	\begin{center}
 		\begin{tabular}{cc}
	 		\includegraphics[scale=0.5]{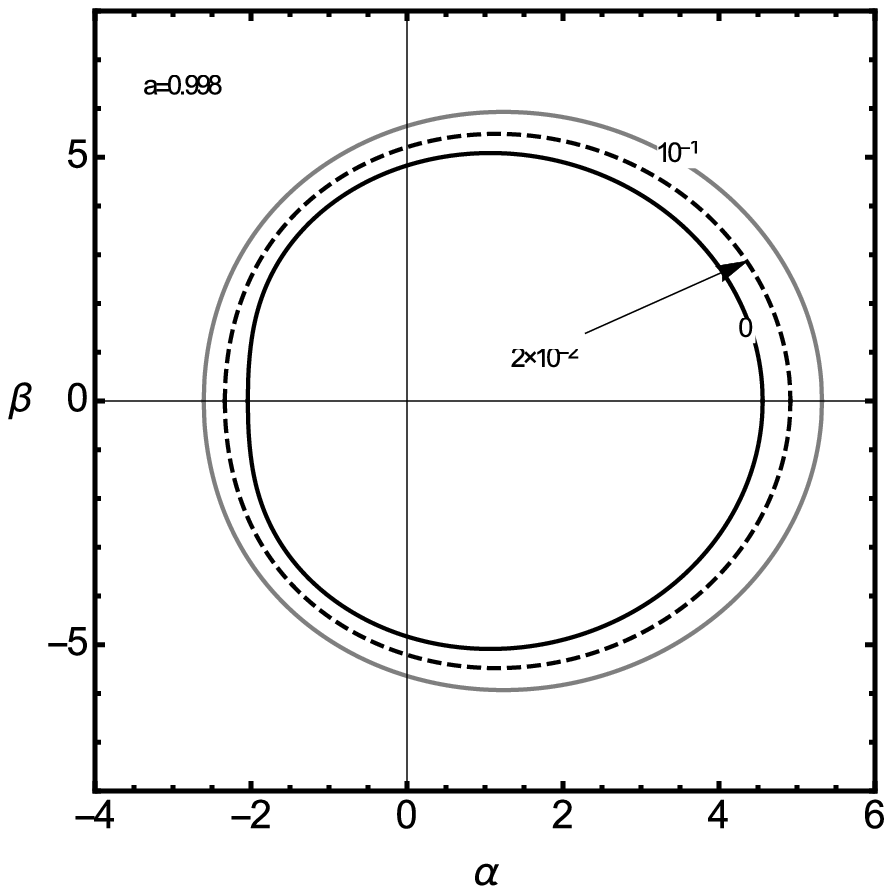} & \includegraphics[scale=0.7]{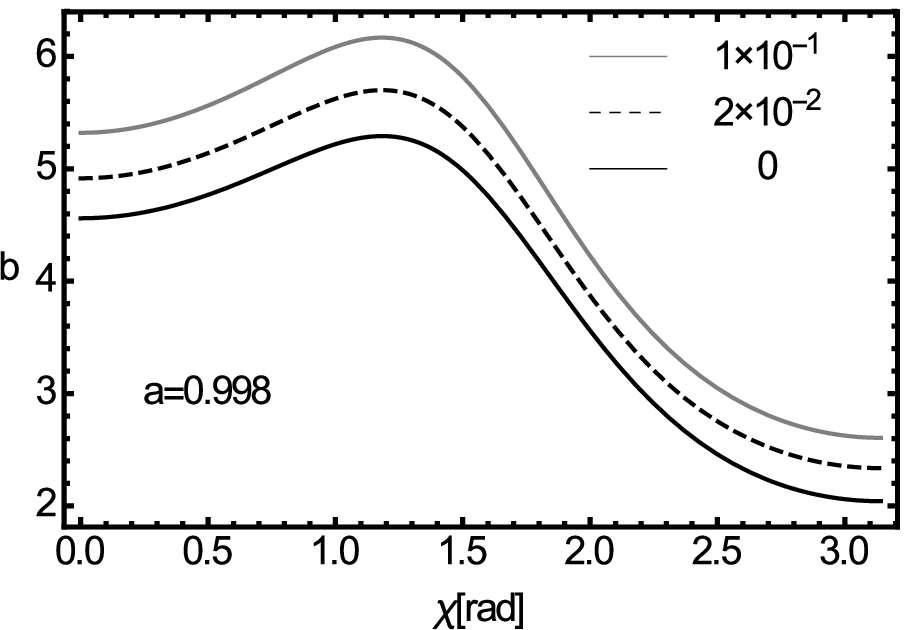}.
 		\end{tabular}
 		\caption{\label{silhouette}Left: Silhouettes of black holes with spin parameter $a=0.998$ and representative values of $c=0$, $2\times 10^{-2}$, and $10^{-1}$. Right: The effect on the silhouette shape given by comparison with the Kerr case by using the quantity $b(\chi;a,c)$. }
 	\end{center}
 \end{figure}

The resulting rotating black hole silhouettes are given for characteristic values of the quintessential field parameter $c=0$, $2\times 10^{-2}$, and $c=10^{-1}$ in Fig. \ref{silhouette}, for the so called canonic value of the black hole spin \cite{Tho:1974:ApJ:}, and large inclination angle $\theta_{o} = 85^{o}$ enabling to visualize strong effect of the black hole rotation. The quintessence field alters the shape of the silhouette by magnifying the shape keeping it almost homogeneous in dependence on the angle $\chi$ as shown in the figure. The magnification increases with increasing quintessential field parameter $c$. 

\section{Spectral line profile}

Finally, we study how the $c$ parameter of the special quintessential rotating black hole modifies the profiles of spectral lines emitted from the innermost parts of the Keplerian disks, $r\in[r_{ms},20M]$.

\begin{figure}[H]
	\begin{center}
	\begin{tabular}{ccc}
		\includegraphics[scale=0.4]{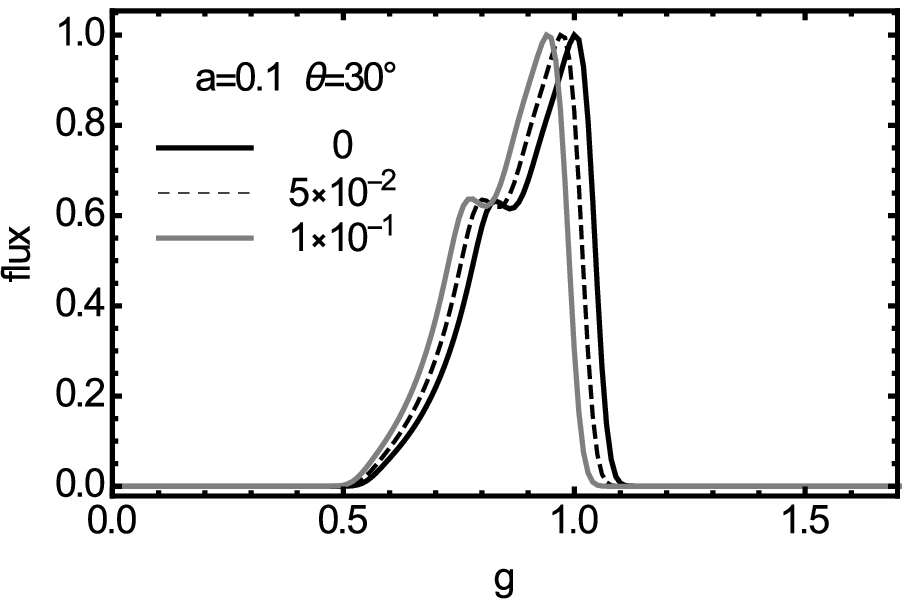}&\includegraphics[scale=0.4]{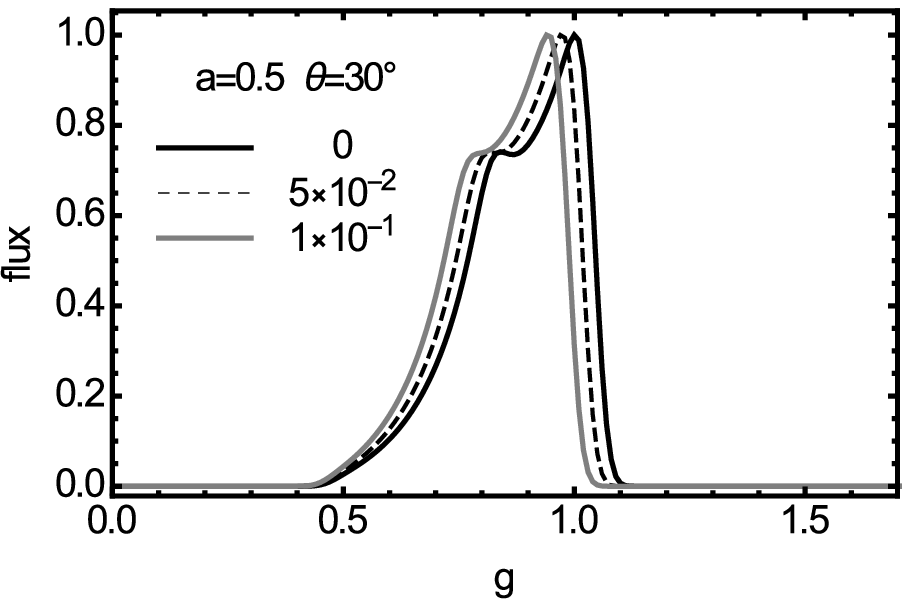}&\includegraphics[scale=0.4]{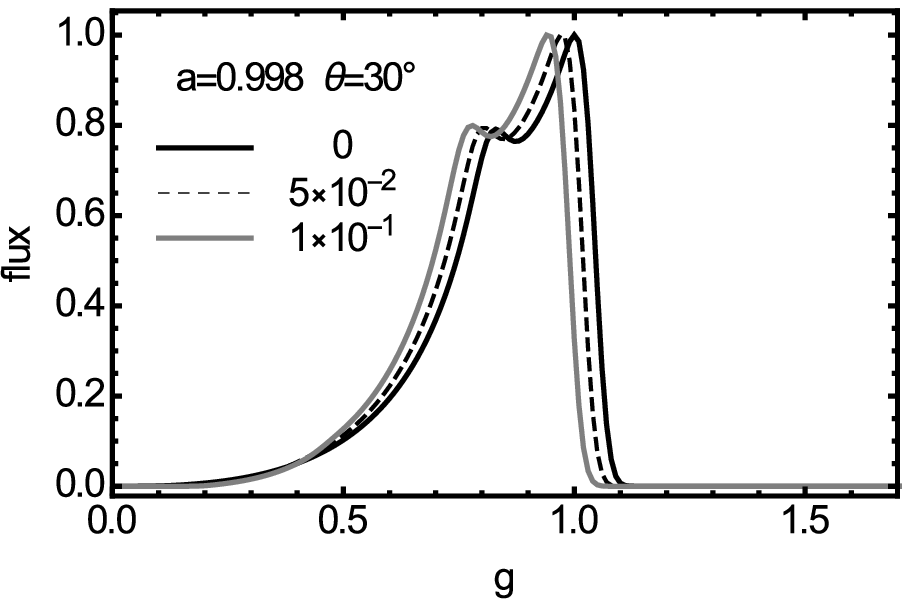}\\
		\includegraphics[scale=0.4]{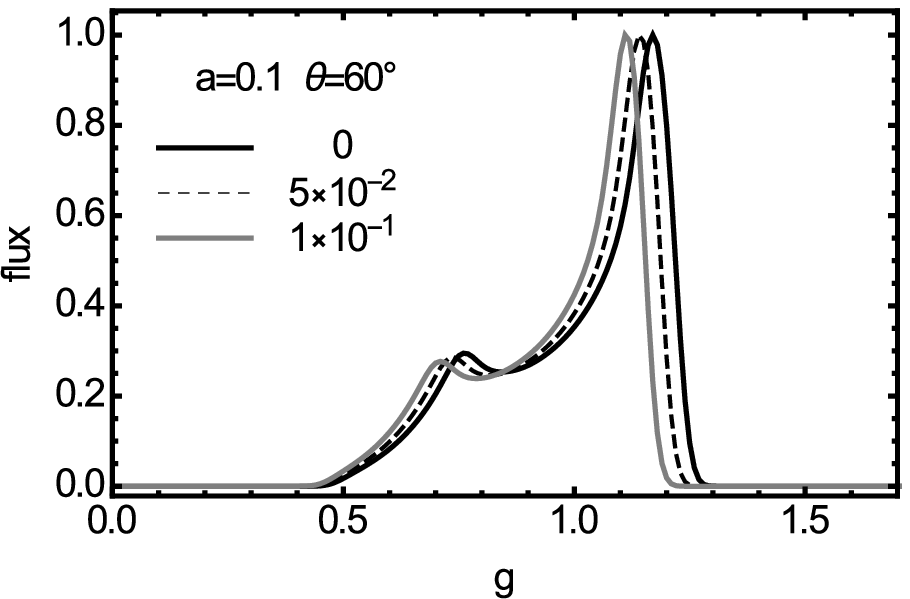}&\includegraphics[scale=0.4]{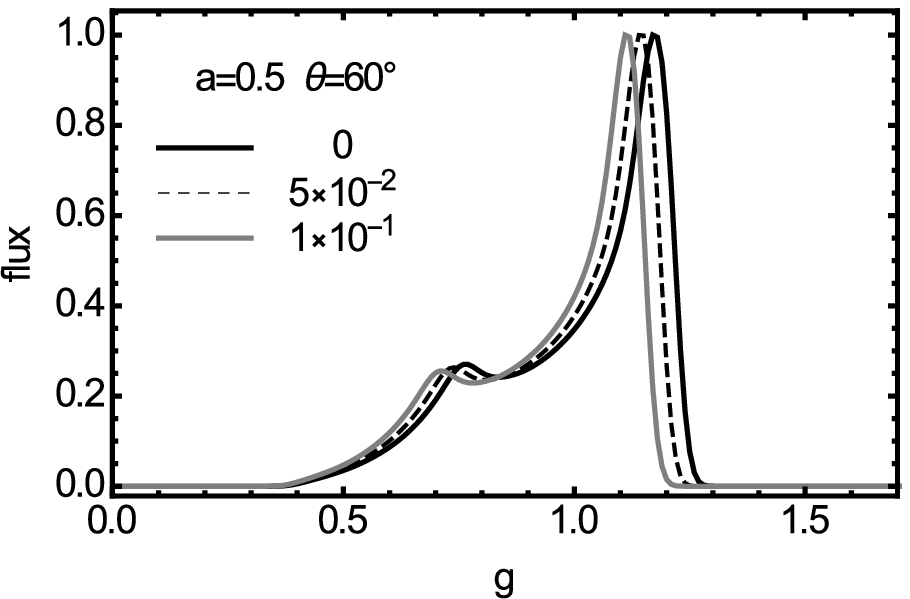}&\includegraphics[scale=0.4]{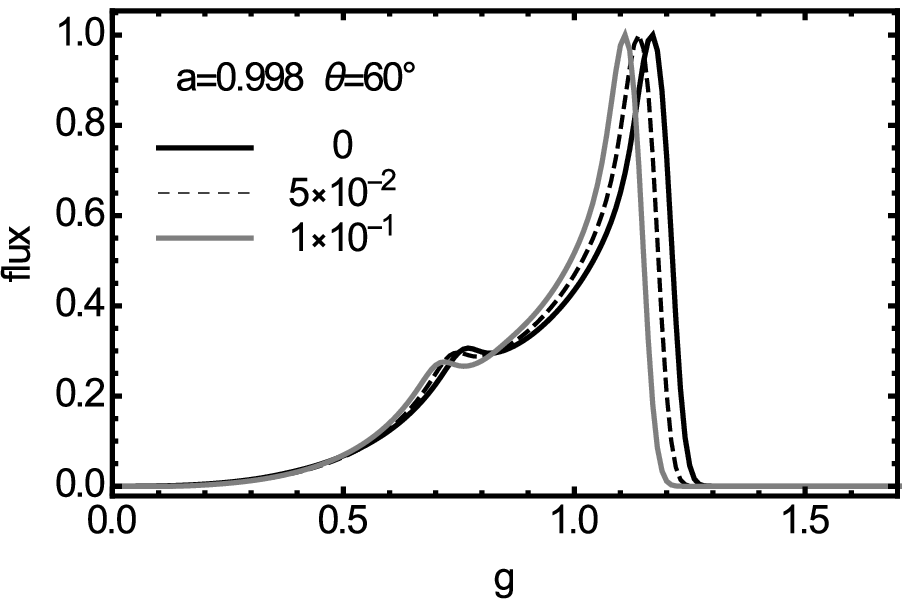}\\
		\includegraphics[scale=0.4]{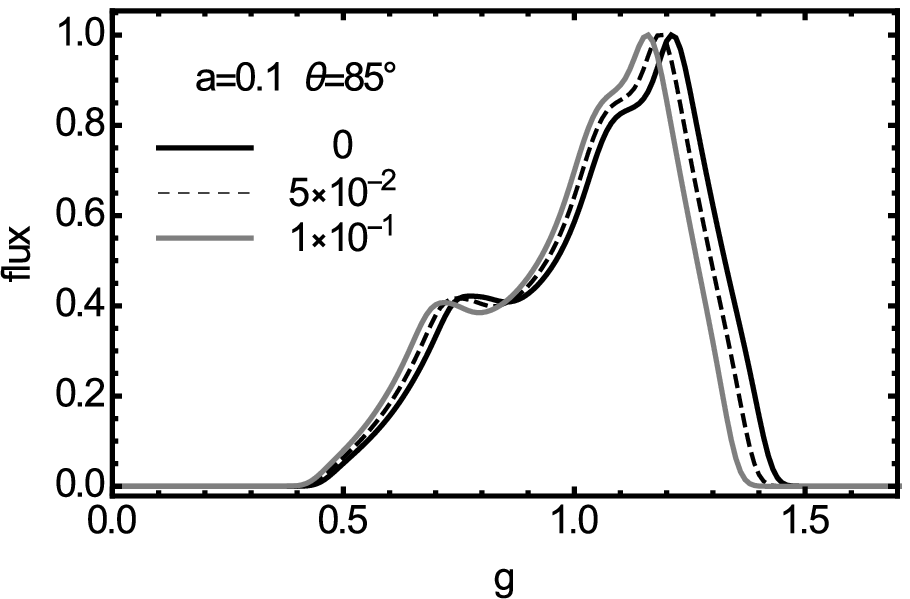}&\includegraphics[scale=0.4]{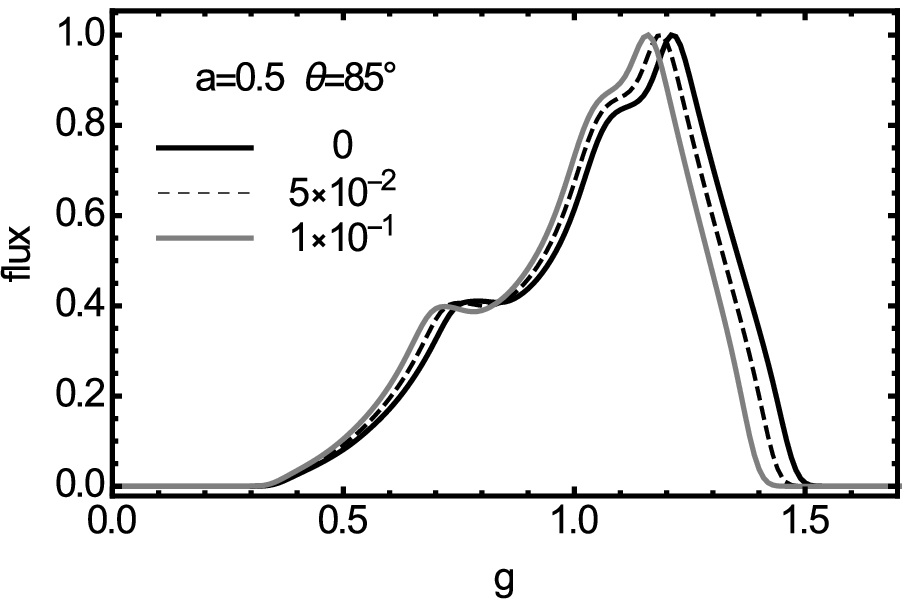}&\includegraphics[scale=0.4]{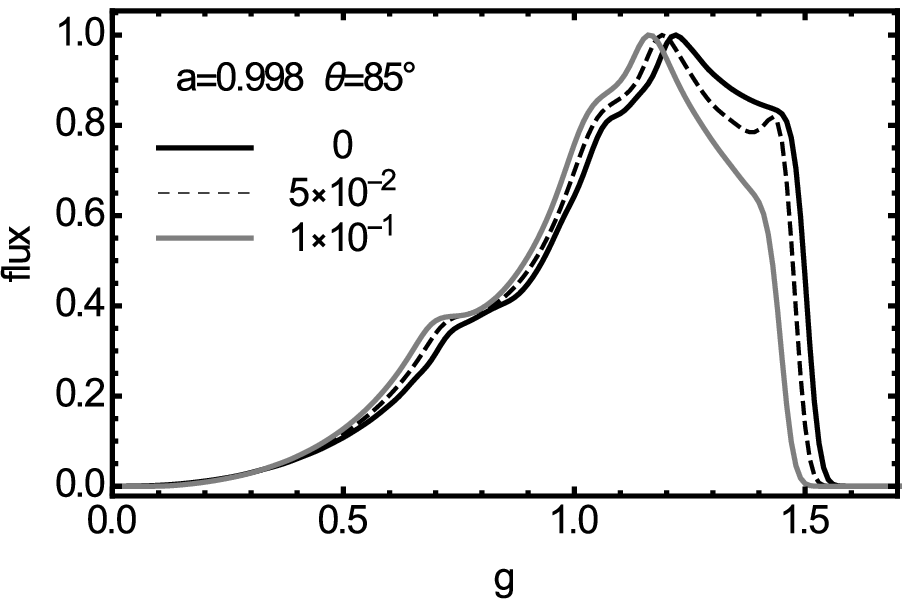}
	\end{tabular}
	\caption{\label{pl}The profiles of spectral lines constructed for two spin parameters $a=0.1$, $0.5$, $0.998$ and three representative values of parameter $c=0$, $5\times 10^{-2}$, and $10^{-1}$. The inclination of the observer is $\theta_o=30^\circ$, $60^\circ$, and $85^\circ$. The disk region that radiates the emission line spans between radii $r_{ISCO}$ and $20M$.}
	\end{center}	
\end{figure}

The profiled spectral line is constructed in the standard way, see, e.g., \cite{Sche-Stu:2009:GRG:}. First, the equations of motion are integrated for a given pair of impact parameters $(\alpha,\beta)$ from the position of the detector to equatorial plane. If the orbits intersects the $\theta=\pi/2$ plane in the region $r_{ms}\leq 20M$, the corresponding frequency shift $g$ is calculated and stored. Second, the photons contributing to the total specific flux are binned by the amount of energy flux related to the frequency shift $g$ and the element of the volume angle related to the detector $\Delta\Pi$ 
\begin{equation}
	\Delta F_i(g)=g^4 r_i^{-p}\Delta\Pi,
\end{equation} 
where the power of the disk radius $p$ denotes the emissivity power-law index. Here we set its value to $p=2$, the frequency shift $g$ is determined by the formula
\begin{equation}
	g=\frac{\sqrt{1-2\rho/r+4a\rho/r \Omega-(r^2+a^2+2a^2\rho/r)\Omega^2}}{1-\Omega l},
\end{equation}
where the Keplerian angular frequency $\Omega=d\phi/dt$, related to the distant observers, reads
\begin{equation}
	\Omega=\frac{2a\rho + (r-2\rho)l_c}{r(r^2+a^2)+2\rho a^2-2a\rho l_c}.
\end{equation}
The corresponding impact parameter $l_c$ related to the circular orbit at radius $r_i$ is determined by the formula
\begin{equation}
	l_c=\frac{L_c}{E_c}=\frac{\sqrt{2}a r^2\Delta A^{1/2}+a^2(B+r^3\Delta')}{a B} ,
\end{equation} 
where 
\begin{eqnarray}
	A&=&2a^2-2\Delta+r\Delta',\\
	B&=&2(a^2-\Delta)\Delta+a^2r\Delta'.
\end{eqnarray}
We run the simulations for three representative spin parameters $a=0.1$, $0.5$, and $0.998$ and three values of the quintessential field parameter $c=0$, $2\times 10^{-2}$, and $10^{-1}$. The $c=0$ case corresponds to the standard Kerr spacetime. The resulting profiles are presented in Fig. \ref{pl}

\section{Summary}

Studies of the astrophysical phenomena in the field of black holes (or alternative strong gravity objects), based on the modified geometry only, can give very important basical information, but they have to be considered very carefully, as additional effects related to the origin of the modified gravity, or the source terms hidden behind the dirty Kerr geometry, could significantly alter the predictions based on the purely geometrical considerations. Here we have considered some optical phenomena around the special quintessential Kerr geometry -- the direct influence of the quintessential field on the optical phenomena is irrelevant and it enters the play only throung the resulting geometry. 

The modifications of both the black hole silhouette and the profile of the spectral line are determined by the functions $b(\chi)$ and $Flux(g)$ which reflect the departure of the particular quantities determined in the specific rotating quintessential metric relative to those determined in the Kerr metric. 

We have clearly demonstrated that in the limiting quintessential field with equation-of-state parameter $\omega_{q}=-1/3$, the resulting geometry of rotating black holes leads, in the case of the characteristic values of the quintessential field parameter $c=0$, $2\times 10^{-2}$, and $10^{-1}$, to significant and specific modifications of the silhouette shape and the profile of the spectral lines generated in the innermost parts of Keplerian disks. 

For the black hole silhouette we have found that the parameter $b$ increases with the parameter $c$ increasing, enlarging the silhouette. The enlargement depends on the angle $\chi$ only slightly. 

In the case of the spectral line profiles we obtained significant shift of the profiled lines to the red end, the shift increases with increasing parameter $c$. The shape of the profiled spectral lines remains unchanged for small and mediate inclination angles. Only in the case of black holes with large spin and for Keplerian disks observed under large inclination angles declination of the profiled spectral lines at the blue end is dereasing with increasing parameter $c$. 

Our results demonstrate that the special quintessential rotating black holes have clear impact on the considered optical phenomena extending the assumed variations predicted for the Kerr metric modification introduced in \cite{Gha-Bam:2016:EPJC:}. 

\section*{Acknowledgement}

The authors acknowledge the Albert Einstein Centre for Gravitation and Astrophysics, supported by the Czech Science Foundation grant No 14-37986G. 


\end{document}